\documentclass[12pt]{article}
\usepackage{a4wide}
\usepackage{dsfont}
\usepackage{cite}
\usepackage{graphicx}
\usepackage{amsfonts}
\usepackage{amstext}
\usepackage{amsmath}
\usepackage{amssymb}
\usepackage{amsthm}
\makeatletter
 
 \@addtoreset{equation}{section}
\makeatother
\def\a{\alpha}
\def\b{\beta}
\def\i{\infty}
\def\l{\lambda}
\def\o{\omega}
\def\Ai{\text{Ai}}

\def\Ai{{\rm Ai}}
\def\b{\beta}

\def\i{\infty}

\def\l{\lambda}

\def\o{\omega}

\begin{document}
\title{On the equal time two-point distribution of the one-dimensional KPZ equation by replica}

\author{T. Imamura
\footnote { Research Center for Advanced Science and Technology, 
The University of Tokyo, Japan,
~email: imamura@jamology.rcast.u-tokyo.ac.jp},
T. Sasamoto
\footnote { Department of mathematics and informatics, 
Chiba University, Japan, ~email: sasamoto@math.s.chiba-u.ac.jp}~$^{\ddagger}$,
H. Spohn
\footnote{Zentrum Mathematik, Technische Universit\"at Mu\"nchen,  Germany, 
~email: sasamoto@ma.tum.de, spohn@ma.tum.de}
}

\maketitle

\noindent
\textbf{Abstract}. In a recent contribution, Dotsenko establishes a Fredholm determinant formula for the two-point 
distribution of the KPZ equation in the long time limit and starting from narrow wedge 
initial conditions. We establish that his expression is identical to the Fredholm determinant 
resulting from the Airy$_2$ process.

\section{Introduction}
The Kardar-Parisi-Zhang (KPZ) equation is a stochastic evolution equation for a height function $h(x,t)$,
which models a surface growing through random deposition \cite{KPZ1986}.  
It was noted early on that the $N$-th exponential moment of the height function 
can be expressed through the propagator of $N$ quantum particles interacting with a 
short range attractive potential \cite{Kardar1987}. In one spatial dimension, this would be the attractive 
Lieb-Liniger Hamiltonian. For its propagator we have now rather concise expressions, which 
allows one to carry out sophisticated replica computations so to obtain exact distribution 
functions for the height \cite{Dotsenko2010,CLDR2010,CLD2011,CLD2012,IS2011a,IS2012,IS2013,PS2011,PS2011a}. 
The various cases are distinguished by the initial conditions
(narrow wedge, flat, two-sided Brownian motion in space variable $x$), by the location of the space-time reference points, 
and by possibly taking a scaling limit (\textit{e.g.} $t\to\i$). 

In the recent contribution  \cite{Dotsenko2013p}, Dotsenko develops a novel summation formula, which he uses
to study narrow wedge initial conditions. In \cite{Dotsenko2013pa} he considers jointly 
$h(0,t),h(0,2t)$, while in \cite{Dotsenko2013p} he considers jointly
$h(x_1,t),h(x_2,t)$, both in the scaling limit $t\to\infty$. It is this latter case which is of interest to us. 
Dotsenko obtains  a Fredholm determinant for the joint distribution.  
Based on universality, Dotsenko's
expression should agree with the corresponding Fredholm determinant of the Airy$_2$ process, 
but merely a direct inspection does not suffice to confirm the conjecture. 

In our note we close the gap and prove the desired identity. As a by-product,
the factorization assumption in 
\cite{PS2011} can be understood  \textit{a posteriori}  as omitting the ``cross terms" of Dotsenko's summation formula.
\section{One-point distribution}
The notations in \cite{Dotsenko2013p,PS2011} differ from each other. To make 
our contribution self-contained and to fix the notation, we briefly recall the replica 
computation for the one-point distribution. 

In dimensionless form the KPZ equation reads
\begin{equation}
 \frac{\partial}{\partial t} h 
 = 
 \frac12 \left( \frac{\partial h}{\partial x} \right)^2
 +
 \frac12 \frac{\partial^2 h}{\partial x^2}
 +
 \eta\,,
\label{KPZ}
\end{equation}
where $h=h(x,t)$ is the height function at location $x \in \mathbb{R}$ and time $t\geq 0$ and $\eta$ is normalized Gaussian white noise.
Through the Cole-Hopf transformation, $Z=\mathrm{e}^h$, 
(\ref{KPZ}) turns into the stochastic heat equation, 
\begin{equation}
 \frac{\partial}{\partial t} Z
 = 
  \frac12 \frac{\partial^2 Z}{\partial x^2}
 +
 \eta Z \,.
\end{equation} 
We will consider only the narrow wedge initial condition $Z(x,0)=\delta(x)$. 

Let us set 
\begin{equation}
 h(x,t)
 = 
 -\tfrac{1}{24}t
 +
 \gamma_t \tilde{h}(x,t)\,,
\end{equation} 
$\gamma_t=(t/2)^{1/3}$, and introduce the generating function
\begin{equation}
 G_{t,x}(s)  = \big\langle \exp\big( -\mathrm{e}^{\gamma_t \tilde{h}(x,t)-s} \big) \big\rangle
                   = \big\langle \exp\big( -\mathrm{e}^{(t/24) - s} Z(x,t) \big) \big\rangle \,.
\end{equation}
Since $\lim_{t\to \infty} \exp( -\mathrm{e}^{-\gamma_t x}) = \theta(x)$, $ \theta$ the step function,
$\theta(x)=1$ for $x>0$, $\theta(x)=0$ for $x<0$, it holds 
\begin{equation}
 \lim_{t\to\infty} G_{t,2\gamma_t^2 x}(\gamma_t (s-x^2))
 =  \lim_{t\to\infty} \big\langle \theta (s-\tilde{h}(2\gamma_t^2 x,t) - x^2) \big\rangle 
 =  \lim_{t\to\infty} \mathrm{Prob} \big[ \tilde{h}(2\gamma_t^2 x,t) +x^2 \leq s \big] \,. 
  \end{equation}
$G_{t,x}(s)$ is a moment generating function for
$Z(x,t)$,  
\begin{equation}
 G_{t,x}(s) = \sum_{N=0}^{\infty} \frac{(-\mathrm{e}^{- s})^N}{N!} \big\langle Z(x,t)^N \big\rangle \mathrm{e}^{tN/24} \,.
\end{equation}
The $N$-th moment, $\langle Z(x,t)^N \rangle$,  can be written using the eigenvalues and  
eigenfunctions of the $\delta$-Bose gas with $N$ particles,  
\begin{equation}
 \langle Z(x,t)^N \rangle 
 = 
 \langle x| \mathrm{e}^{-H_N t}|0\rangle 
 =
 \sum_z \langle x|\Psi_z\rangle \langle \Psi_z|0\rangle \mathrm{e}^{-E_z t} \,,
\end{equation}
where $|x\rangle= |x,\ldots , x\rangle$, $N$-times, $H_N$ is the Hamiltonian of the 
$\delta$-Bose gas of $N$ particles, and 
$z$ labels the eigenvalues $E_z$ and eigenstates $|\Psi_z\rangle$.  
Using the usual parametrization of the eigenstates \cite{Dotsenko2010,IS2011a}, 
for fixed $N$ one writes
\begin{equation}
 E_z 
 = 
 \frac12 \sum_{j=1}^N z_j^2
 =
 \frac12 \sum_{\alpha=1}^M n_{\alpha} q_{\alpha}^2
 -
 \frac{1}{24}\sum_{\alpha=1}^M (n_{\alpha}^3-n_{\alpha})\,,
\end{equation}
\begin{equation}
 \langle x|\Psi_z\rangle \langle \Psi_z|0\rangle
 =
  |\langle 0|\Psi_z\rangle |^2 \mathrm{e}^{\mathrm{i} \sum_{\alpha=1}^M n_{\alpha} q_{\alpha} x}\,,
\end{equation}
\begin{equation}
  |\langle 0|\Psi_z\rangle |^2
  =
  N! \det\left( \frac{1}{\frac12 (n_{\alpha}+n_{\beta})+\mathrm{i}(q_\a-q_\b)} \right)_{\a,\b=1}^M\,,
\end{equation}
\begin{equation}
\sum_z 
= 
\sum_{M=1}^N \frac{1}{M!} \prod_{\a=1}^M \sum_{n_\a=1}^\i 
\int_{-\i}^\i \frac{dq_\a}{2\pi} \delta_{\sum_{\a=1}^M n_\a,N}\,.
\end{equation}
Hence
\begin{eqnarray}
 &&\hspace{-30pt}G_{t,x}(s) = \sum_{M=0}^\i \frac{(-1)^M}{M!} \prod_{\a=1}^M 
 \int_{-\i}^\i \frac{dq_\a}{2\pi} \sum_{n_\a=1}^\i (-1)^{n_\a-1}
 \mathrm{e}^{n_\a^3t/24 + n_\a(\mathrm{i} q_\a x-\frac12 q_\a^2 t - s)} \notag\\
 &&\hspace{100pt} \times
 \det\left( \frac{1}{\frac12 (n_{\alpha}+n_{\beta})+\mathrm{i}(q_\a-q_\b)} \right)_{\a,\b=1}^M
\end{eqnarray}
with the understanding that the term with $M=0$ equals 1. 

Using 
\begin{equation}
 \frac{1}{\frac12 (n_{\alpha}+n_{\beta})+\mathrm{i}(q_\a-q_\b)}
 =
  \int_0^\i d\o_\a \mathrm{e}^{-( \frac12 (n_{\alpha}+n_{\beta})+\mathrm{i}(q_\a-q_\b)  )\o_\a} 
\end{equation}
and a simple identity for determinants, one has 
\begin{eqnarray}
 &&\hspace{-40pt}G_{t,x}(s-x^2/(2t)) = \sum_{M=0}^\i \frac{(-1)^M}{M!} \prod_{\a=1}^M \int_0^\i d\o_\a 
 \int_{-\i}^\i \frac{dq_\a}{2\pi} \sum_{n_\a=1}^\i (-1)^{n_\a-1}
  \notag\\
 &&\hspace{10pt}  \times \mathrm{e}^{n_\a^3t/24 + n_\a(\mathrm{i} q_\a x-\frac12 q_\a^2 t- s + x^2/2t)}
 \det\left(  \mathrm{e}^{-\frac12 n_{\alpha}(\o_\a+\o_\b) - \mathrm{i} q_\a(\o_\a-\o_\b) }  \right)_{\a,\b=1}^M \,.
\end{eqnarray}

We rescale $s\to \gamma_t s, x\to 2\gamma_t^2 x, 
\o_\a \to \gamma_t \o_\a, q_\a \to q_\a / \gamma_t$.
Shifting $q_\a \to q_\a + \mathrm{i} x$ and noting that $\mathrm{e}^{x(\o_\a-\o_\b)}$ 
inserted in the determinant does not change its value, we arrive at  
\begin{equation}
 G_{t,2\gamma_t^2 x}(\gamma_t (s-x^2)) = \det(1-K_{t,s})\,,
\end{equation}
where the kernel of $K_{t,s}$ is given by 
\begin{equation}
 K_{t,s}(\o,\o')
 =
 \int_{-\i}^\i \frac{dq}{2\pi} \mathrm{e}^{\mathrm{i}q(\o'-\o)}
 \sum_{n=1}^\i (-1)^{n-1} \mathrm{e}^{n^3t/24 -\gamma_t n\left( q^2+\frac12(\o+\o')-s \right)} .
\end{equation}
Using 
\begin{equation}
 \mathrm{e}^{\l^3 n^3/3} = \int_{-\i}^\i dy \Ai (y) \mathrm{e}^{\l n y} 
\end{equation}
with $\l = t^{1/3}/2 = \gamma_t/2^{2/3}$ and shifting $y\to y+q^2+\frac12(\o+\o')$, one obtains
\begin{eqnarray}
 &&\hspace{-20pt} K_{t,s}(\o,\o')
 =
 2^{2/3}\int_{-\i}^\i \frac{dq}{2\pi} \mathrm{e}^{\mathrm{i}q(\o'-\o)}
  \int dy\Ai \big( 2^{2/3}(y+q^2+\tfrac12(\o+\o') \big)\nonumber\\  
&&\hspace{120pt}\times \sum_{n=1}^\i (-1)^{n-1} \mathrm{e}^{-\gamma_t n(s-y)} \,.
\end{eqnarray}
Summing over $n$ and using the identity 
\begin{equation}
 2^{2/3}  \int_{-\i}^\i \frac{dq}{2\pi} \mathrm{e}^{2\mathrm{i}q y} \Ai \big(2^{2/3}(q^2+x)\big)
 =
 \Ai(x+y) \Ai(x-y) \,,
 \label{Ai2AiAi}
\end{equation}
one obtains
\begin{equation}
 K_{t,s}(\o,\o')
 =
 \int_{-\i}^\i dy \Ai(\o+y) \Ai(\o'+y) \frac{\mathrm{e}^{\gamma_t(y-s)}}{1+\mathrm{e}^{\gamma_t(y-s)}} \,.
\end{equation}
In the $t\to\i$ limit, 
\begin{equation}
 \frac{\mathrm{e}^{\gamma_t(y-s)}}{1+\mathrm{e}^{\gamma_t(y-s)}} \to \theta(y-s)
 \label{th_lim}
\end{equation}
and the kernel  of $K_{t,s}$ turns into the kernel for the GUE Tracy-Widom distribution. 
\section{Two-point distribution and factorization assumption}
Next we study the joint distribution of $\tilde{h}( x_1,t ), \tilde{h}( x_2,t )$
for $x_2-x_1=\mathcal{O}(t^{2/3})$  in the limit $t\to\i$. 
Following the strategy of Section 2, we consider the generating function 
\begin{eqnarray}
 &&\hspace{-44pt}G_{t,x_1,x_2}(s_1,s_2) 
 = 
 \big\langle \exp \big( -\mathrm{e}^{\gamma_t \tilde{h}(x_1,t)-s_1} 
                      -\mathrm{e}^{\gamma_t \tilde{h}(x_2,t)-s_2} \big) \big\rangle \notag\\[1ex]
 &&\hspace{30pt}= 
 \big\langle \exp \big( -\mathrm{e}^{(t/24)-  s_1} Z(x_1,t) 
                      -\mathrm{e}^{(t/24) -s_2} Z(x_2,t)  
 \big) \big\rangle\,.
\end{eqnarray}
It holds  
\begin{eqnarray}
&&\hspace{-28pt}\lim_{t\to\infty} G_{t,2\gamma_t^2 x_1,2\gamma_t^2 x_2}(\gamma_t (s_1-x_1^2),
\gamma_t (s_2-x_2^2))\nonumber\\ 
&&\hspace{-8pt}=
\lim_{t\to\infty} \mathrm{Prob} \big[ \tilde{h}( x_1,t ) +x_1^2 \leq s_1, 
                            \tilde{h}( x_2,t ) + x_2^2 \leq s_2 \big] \,.
\end{eqnarray}
By expanding in $\mathrm{e}^{-s_1}$, $\mathrm{e}^{-s_2}$, 
\begin{equation}\label{3.3}
 G_{t,x_1,x_2}(s_1,s_2) 
 = 
 \sum_{L=0}^{\infty} \sum_{R=0}^{\infty} 
 \frac{(-\mathrm{e}^{- s_1})^L}{L!}  \frac{(-\mathrm{e}^{-s_2})^R}{R!}
 \big\langle Z(x_1,t)^L Z(x_2,t)^R \big\rangle \mathrm{e}^{ (L+R)t/24 }\,,
\end{equation}
where the term with $L=0= R$ equals 1. 
Expanding in the eigenstates of the $\delta$-Bose gas one arrives at
\begin{equation}
 \big\langle Z( x_1,t )^L Z ( x_2,t)^R \big\rangle
 =
 \sum_z  |\langle 0|\Psi_z\rangle |^2 \mathrm{e}^{-E_z t} 
 \frac{\langle x_1,x_2;L,R|\Psi_z\rangle}{\langle 0|\Psi_z\rangle }\,,
\end{equation}
where $|x_1,x_2;L,R\rangle = |x_1,\ldots, x_1,
x_2,\ldots, x_2 \rangle$ with $L$ times the argument $x_1$ and $R$ times the argument
$x_2$.

By Dotsenko's summation formula, the last factor in the summation can be written as
\begin{eqnarray}
&&\hspace{-8pt} \frac{\langle x_1,x_2; L,R |\Psi_z\rangle}{\langle 0|\Psi_z\rangle } 
 \delta_{\sum_{\a=1}^M n_\a,L+R} 
 =
 \binom{L+R}{L}^{-1}\sum_{m_\a+l_\a>0} \prod_{\a=1}^M \binom{n_\a}{m_\a} 
 \notag\\
 &&\hspace{-0pt}\times\mathrm{e}^{\mathrm{i}x_1m_\a q_\a+ \mathrm{i}x_2l_\a q_\a- m_\a l_\a x/2} 
 \delta_{m_\a+l_\a,n_\a} \delta_{\sum_{\a=1}^M m_\a,L} \delta_{\sum_{\a=1}^M l_\a,R} 
 \times CT\,,
\end{eqnarray}
where $x=x_2-x_1$ and the summation over $m_\a,l_\a$ is for all nonnegative values except for 
$m_\a=0 = l_\a$. The factor $CT$ denotes the cross term which involves  a product over all 
pairs $\a,\b$, explicitly given by  (\textit{cf}. (26) in \cite{Dotsenko2013p})
\begin{eqnarray}
&&\hspace{-20pt}CT = \\
&&\hspace{-20pt}\prod_{\a\neq \b}
\frac{\Gamma[1+\frac12(m_\a+m_\b+l_\a+l_\b)+\mathrm{i}(q_\a-q_\b)]
        \Gamma[1+\frac12(-m_\a+m_\b+l_\a-l_\b)+\mathrm{i}(q_\a-q_\b)]}
       {\Gamma[1+\frac12(-m_\a+m_\b+l_\a+l_\b)+\mathrm{i}(q_\a-q_\b)]
        \Gamma[1+\frac12(m_\a+m_\b+l_\a-l_\b)+\mathrm{i}(q_\a-q_\b)]}\,. \nonumber
\end{eqnarray}
Substituting in (\ref{3.3}), we obtain
\begin{eqnarray}
 &&\hspace{-40pt}G_{t,x_1,x_2}(s_1,s_2) 
  = 
  \sum_{M=0}^\i \frac{(-1)^M}{M!} \prod_{\a=1}^M 
 \int_{-\i}^\i \frac{dq_\a}{2\pi} \sum_{m_\a+l_\a>0} (-1)^{m_\a+l_\a-1} \binom{m_\a+l_\a}{m_\a}
 \notag\\
 &&\hspace{40pt}\times\det\left( \frac{1}{\frac12(m_\a+l_\a)+\frac12(m_\b+l_\b)+\mathrm{i}(q_\a-q_\b)} \right)_{\a,\b=1}^M
 \notag\\[1ex]
 &&\hspace{40pt}\times 
 \mathrm{e}^{(m_\a+l_\a)^3t/24 -\frac12 (m_\a+l_\a)q_\a^2 t  -m_\a s_1- l_\a s_2 
       +\mathrm{i}x_1m_\a q_\a+ \mathrm{i}x_2l_\a q_\a- m_\a l_\a x/2} \times CT\,.
\label{Grep}
\end{eqnarray}
On the other hand, the expression (4.8) in \cite{PS2011} after factorization reads, upon replacing 
$x \to x/2$ there to match with our notation\footnote{Note that the definitions of $x$ in \cite{Dotsenko2013p} and \cite{PS2011} differ by factor of 2. Also 
1/2 in (3.21),(3.23),(3.27) of \cite{PS2011} should rather read 1/4.},  
\begin{eqnarray}
 &&\hspace{-20pt}G_{t,x_1,x_2}^{\#}(s_1,s_2) 
  = 
 \sum_{M=0}^\i \frac{(-1)^M}{M!} \prod_{\a=1}^M 
 \int_{-\i}^\i \frac{dq_\a}{2\pi} \sum_{n_\a=1}^{\i} (-1)^{n_\a-1} 
 \det\left( \frac{1}{\frac12 n_\a +\frac12 n_\b+\mathrm{i}(q_\a-q_\b)} \right)_{\a,\b=1}^M
 \notag\\
 &&\hspace{50pt}\times 
 \prod_{\a=1}^M \mathrm{e}^{n_\a^3t/24 -\frac12 n_\a q_\a^2 t } 
  \mathrm{e}^{-\frac12 x \partial_1 \partial_2} 
  \left( \mathrm{e}^{\mathrm{i}x_1 q_\a- s_1}+\mathrm{e}^{\mathrm{i}x_2 q_\a- s_2} \right)^{n_\a}\, ,
  \label{Gfact}
\end{eqnarray}
where $\partial_i = \partial_{s_i},i=1,2$. 
Binomially expanding the last factor on the RHS, we note that
\begin{eqnarray}
 &&\hspace{-30pt} \mathrm{e}^{-\frac{1}{2} x \partial_1 \partial_2} 
  \left( \mathrm{e}^{ix_1 q_\a - s_1}+\mathrm{e}^{\mathrm{i}x_2 q_\a- s_2} \right)^{n_\a}
 \notag\\
  &&\hspace{-10pt}=
  \sum_{m_\a=0}^{n_\a}   \mathrm{e}^{-\frac12 x m_\a(n_\a-m_\a)}
  \binom{n_\a}{m_\a} \mathrm{e}^{m_\a(\mathrm{i} x_1 q_\a - s_1)}
  \mathrm{e}^{(n_\a-m_\a)(\mathrm{i} x_2 q_\a - s_2)}\,.
\end{eqnarray}
Rewriting the summation over $n_\a$ in (\ref{Gfact}) by setting $l_\a=n_\a-m_\a$,  this expression is 
the same as (\ref{Grep}) except for the last factor $CT$. 
Hence the factorization assumption in \cite{PS2011} amounts to set the cross term $
CT=1$. 
\section{Dotsenko's formula yields the Airy$_2$ process}
In \cite{PS2011} it is established that under factorization the limit $t\to\i$ of the expression 
in the last section yields indeed 
the two-point distribution of the Airy$_2$ process. On the other hand in \cite{Dotsenko2013p}
Dotsenko argues that, after a suitable analytic continuation, $CT\to 1$ as $t\to\i$ and then 
writes a Fredholm determinant expression for the limiting two-point distribution. 
One expects that Dotsenko's formula should coincide with the two-point function of  the Airy$_2$ process.
In this section  we establish rather straightforwardly the equivalence between the expression (51)
in \cite{Dotsenko2013p}  and the $t\to\i$ limit of (4.19) in \cite{PS2011}. 

Substituting 
$s_i \to \gamma_t s_i$, $x_i\to  2 \gamma_t^2 x_i$, $i=1,2$,  \textit{cf}. \cite{PS2011} Eq. (2.31), 
(4.19) in \cite{PS2011} turns into 
\begin{equation}
 G_{t,2\gamma_t^2 x_1,2\gamma_t^2 x_2}^{\#}(\gamma_t (s_1-x_1^2),\gamma_t (s_2-x_2^2)) 
 =
 \det(1-N)\,,
\end{equation}
where the kernel of $N$ is given by
\begin{eqnarray}
&&\hspace{-10pt}N(z_1,z_2)
=
\mathds{1}_{\{z_1,z_2>0\}} \int_{-\i}^\i du \mathrm{e}^{- 2x\partial_1\partial_2}
\mathrm{e}^{-(x_1\partial_1+ x_2\partial_2)(\partial_{z_1}-\partial_{z_2})}
\Phi\big(\gamma_t(u-s_1),\gamma_t(u-s_2)\big) \notag\\ 
&&\hspace{50pt} \times \Ai(z_1+u)\Ai(z_2+u)\,.
\label{kernelN}
\end{eqnarray}
Here 
\begin{equation}
\Phi(u,v) =\frac{\mathrm{e}^u+\mathrm{e}^v}{\mathrm{e}^u+\mathrm{e}^v+1}
= 
\sum_{l+m>0}(-1)^{l+m-1} \binom{l+m}{m} \mathrm{e}^{mu+lv}\, .
\label{Phi2}
\end{equation}
Therefore
\begin{eqnarray}
 &&\hspace{-40pt}\lim_{t\to\i} \Phi \big(\gamma_t(u-s_1),\gamma_t (u-s_2)\big)
 \notag\\
 &&\hspace{10pt}=\theta(u-s_1)+\theta(u-s_2) -\theta(u-s_1)\theta(u-s_2)
 =:
 \Phi_\i(s_1,s_2,u)
 \label{Phi2lim}
\end{eqnarray}
and hence 
\begin{equation}
\lim_{t\to\i} \mathrm{Prob} 
\big[ \tilde{h}(2 \gamma_t^2 x_1,t)+x_1^2 \leq s_1, \tilde{h}(2\gamma_t^2 x_2,t)+x_2^2\leq s_2 \big]
=\det(1-N_\i)\,,
\label{4.5}
\end{equation}
where
\begin{equation}
N_\i(z_1,z_2)
=
\mathds{1}_{\{z_1,z_2>0\}} \int_{-\i}^\i du \mathrm{e}^{- x\partial_1\partial_2}
\mathrm{e}^{ -(x_1\partial_1+x_2\partial_2)(\partial_{z_1}-\partial_{z_2})}
\Phi_\i(s_1,s_2,u) 
\Ai(z_1+u)\Ai(z_2+u)\, .
\label{Ninf}
\end{equation}
The derivative with respect to $s_1$, $s_2$ may seem rather formal,
since it is acting on a step function. However, in later sections of \cite{PS2011} there are equivalent expressions for 
$N_\i$, for which such difficulties do not arise, compare with Section 5.

Turning to Eq. (51) in \cite{Dotsenko2013p}, with the understanding 
that one should replace $x \to 2^{4/3} x$, $s \to  2^{2/3} s$ 
to conform with the notations up to now, the probability in (\ref{4.5}) is written as a Fredholm determinant with 
operator kernel 
\begin{eqnarray}
 &&\hspace{-20pt}A(\o_1,\o_2)\notag\\
 &&\hspace{0pt}=
 \int_0^\i dy \int_{-\i}^\i \frac{dq}{2\pi} 
 \Ai \big(y+q^2+s_1+\o_1+\o_2-\mathrm{i}qx_1 \big) \mathrm{e}^{\mathrm{i}q(\o_2-\o_1)} \notag\\
 &&\hspace{20pt}+
 \int_0^\i dy \int_{-\i}^\i \frac{dq}{2\pi} 
 \Ai\left( y+q^2+s_2+\o_1+\o_2+ \mathrm{i}qx_2 \right) \mathrm{e}^{\mathrm{i}q(\o_2-\o_1)} \notag\\
 &&\hspace{20pt} -
 \int_{-\i}^\i dy \int_{-\i}^\i \frac{dq}{2\pi}\, (2\pi)^{-3/2}\!\!\int d\xi_1 d\xi_2 d\xi_3
 \Ai\big(y+q^2+\o_1+\o_2-\mathrm{i}\xi_3 \sqrt{x} \,\big) \mathrm{e}^{\mathrm{i}q(\o_2-\o_1)} \notag\\[1ex]
 &&\hspace{40pt} \times\mathrm{e}^{-\frac12(\xi_1-\mathrm{i}qx_1/\sqrt{x})^2 -\frac12(\xi_2-\mathrm{i}qx_2/\sqrt{x})^2 -\frac12 \xi_3^2}
 \theta\big(y-s_1+\xi_1\sqrt{x}\,\big) \theta(y-s_2+\xi_2\sqrt{x}\,\big) \notag\\[1ex]
 &&\hspace{0pt}
 =
 \int_{-\i}^\i \frac{dq}{2\pi} \int dy \Ai \big(y+q^2 +\o_1+\o_2\big) \mathrm{e}^{\mathrm{i}q(\o_2-\o_1)} 
  \big\{  \theta\big( y -s_1+\mathrm{i}qx_1 \big) + \theta \big( y -s_2+\mathrm{i}qx_2 \big)  \notag\\ 
 &&\hspace{20pt} 
 - (2\pi)^{-3/2}\!\!\int d\xi_1 d\xi_2 d\xi_3\mathrm{e}^{-\frac12(\xi_1^2+\xi_2^2+\xi_3^2)}
 \mathrm{e}^{-(\xi_1+\mathrm{i}\xi_3)\sqrt{x}\partial_1}   \mathrm{e}^{-(\xi_2+\mathrm{i}\xi_3)\sqrt{x}\partial_2}  \notag\\ 
 &&\hspace{140pt} \times\theta\big( y -s_1+\mathrm{i}qx_1 \big) \theta\big( y-s_2+\mathrm{i}qx_2 \big)  \big\}\,.
\end{eqnarray}
Here the step functions with complex arguments are to be understood as resulting from a limit in
analogy to (\ref{th_lim}).  
The last term can be rewritten as  
\begin{eqnarray}\label{4.10}
 &&\hspace{-10pt}(2\pi)^{-3/2} \int d\xi_1 d\xi_2 d\xi_3\mathrm{e}^{-\frac12(\xi_1^2+\xi_2^2+\xi_3^2)}
 \mathrm{e}^{-(\xi_1+i\xi_3)\sqrt{x}\partial_1}   \mathrm{e}^{-(\xi_2+\mathrm{i}\xi_3)\sqrt{x}\partial_2}  
 \theta \big( y-s_1+\mathrm{i}qx_1 \big) \notag\\
 &&\hspace{40pt} \times\theta\big( y-s_2+\mathrm{i}qx_2 \big) =
 \mathrm{e}^{-x\partial_1\partial_2} \theta \big( y-s_1+\mathrm{i}qx_1 \big) 
 \theta\big( y-s_2+\mathrm{i}qx_2  \big)\,.
\end{eqnarray}
{\it Proof.} It is easy to see that
\begin{eqnarray}
  &&\hspace{-30pt}\mathrm{e}^{-\frac12 x\partial_1\partial_2}  \mathrm{e}^{-ms_1-ls_2}
  =
   \mathrm{e}^{-ms_1-ls_2-\frac12 xml}
 \notag\\
  &&\hspace{10pt}= 
  (2\pi)^{-3/2}\!\! \int d\xi_1 d\xi_2 d\xi_3
  \mathrm{e}^{-\frac12(\xi_1^2+\xi_2^2+\xi_3^2)+m\sqrt{x}\xi_1
  +l\sqrt{x}\xi_2+i(m+l)\sqrt{x}\xi_3}  \mathrm{e}^{-ms_1-ls_2} \notag\\
  &&\hspace{10pt}=
   (2\pi)^{-3/2}\!\! \int d\xi_1 d\xi_2 d\xi_3 \mathrm{e}^{-\frac12(\xi_1^2+\xi_2^2+\xi_3^2)}
  \mathrm{e}^{-(\xi_1+i\xi_3)\sqrt{x}\partial_1}    \mathrm{e}^{-(\xi_2+i\xi_3)\sqrt{x}\partial_2}   
  \mathrm{e}^{-ms_1-ls_2}\, .
  \label{lin}
\end{eqnarray}
Multiplying both sides by 
$ (-1)^{l+m-1}\mathrm{e}^{mu+lv+\mathrm{i}(m x_1+l x_2)q}\binom{m+l}{m}$,
substituting $s_i \to \gamma_t s_i$, $x_i\to 2\gamma_t^2 x_i$, and summing over $l,m$, we obtain
\begin{eqnarray}
&&\hspace{-30pt} \mathrm{e}^{-x\partial_1\partial_2} \Phi\big(\gamma_t(u-s_1+ \mathrm{i}x_1q), 
\gamma_t(u-s_2+\mathrm{i}x_2q)\big)
\notag\\[1ex]
&&\hspace{-10pt}=
(2\pi)^{-3/2}\!\! \int d\xi_1 d\xi_2 d\xi_3 \mathrm{e}^{-\frac12(\xi_1^2+\xi_2^2+\xi_3^2)} \notag\\
 &&\hspace{2pt}\times 
 \mathrm{e}^{-(\xi_1+\mathrm{i}\xi_3)\sqrt{x}\partial_1}    \mathrm{e}^{-(\xi_2+\mathrm{i}\xi_3)\sqrt{x}\partial_2} 
 \Phi\big(\gamma_t(u-s_1+\mathrm{i}x_1q), \gamma_t(u-s_2+\mathrm{i}x_2q)\big)\,.
\end{eqnarray}
Taking the limit $t\to\i$, we arrive at the desired equality. \medskip\qed

Using Eq. (\ref{4.10}) one writes
\begin{eqnarray}
 &&\hspace{-10pt}A(\o_1,\o_2) 
 =
 \int_{-\i}^\i \frac{dq}{2\pi} \int dy \Ai(y+q^2+\o_1+\o_2) \mathrm{e}^{\mathrm{i}q(\o_2-\o_1)} 
 \notag\\[1ex]
 &&\hspace{50pt} \times\mathrm{e}^{-x\partial_1\partial_2}
  \Phi_\i\big( s_1-\mathrm{i}x_1q, s_2-\mathrm{i}x_2 q,y\big) \notag\\[1ex]
 &&\hspace{20pt}=
 2 \int_{-\i}^\i \frac{dq}{2\pi} 
\mathrm{e}^{\mathrm{i}q 2^{1/3}\big( (\o_2-\o_1)-(x_1 \partial_1+x_2 \partial_2) \big) }  
 \int dy \Ai\big( 2^{2/3}(y+q^2+2^{-2/3}(\o_1+\o_2) \big)  
 \notag\\[1ex]
 &&\hspace{50pt} \times\mathrm{e}^{-x\partial_1\partial_2} \Phi_\i( s_1, s_2,2^{2/3}y) \,.
\end{eqnarray}
We switch to $x,s$ as in (\ref{Ninf}) and use (\ref{Ai2AiAi}) to obtain
\begin{eqnarray}
 &&\hspace{-10pt}A(\o_1,\o_2) 
 =
 2^{1/3} \int dy \Ai \big( y+2^{1/3}\o_1- (x_1\partial_1+ x_2\partial_2) \big)
 \Ai \big( y+2^{1/3}\o_2+(x_1\partial_1+x_2\partial_2) \big)\notag\\
 &&\hspace{50pt}\times \mathrm{e}^{-x\partial_1\partial_2} \Phi_\i( s_1, s_2,y)\, . 
\label{4.12}
\end{eqnarray} 
Since $\Phi_\i$ can be written as a limit of a sum of exponentials, see (\ref{Phi2}),(\ref{Phi2lim}), 
the derivatives $\partial_i$, $i=1,2$, inside the Airy functions can be regarded as numbers.  
Finally one can rewrite (\ref{4.12}) as 
\begin{eqnarray}
&&\hspace{-20pt}A(\o_1,\o_2) 
 =
 2^{1/3} \int dy \mathrm{e}^{-x\partial_1\partial_2}  
 \mathrm{e}^{-2^{-1/3} (x_1\partial_1+x_2 \partial_2)(\partial_{\o_1}- \partial_{\o_2})}
 \nonumber\\
 &&\hspace{40pt}\times\Ai(y+2^{1/3}\o_1) \Ai(y+2^{1/3}\o_2) \Phi_\i( s_1, s_2,y) \,,
\end{eqnarray} 
which is the same as the kernel $N_\i$ of (\ref{Ninf}) with the identification 
$z_i=2^{1/3}\o_i,i=1,2$. 

We draw one conclusion from our computation. Both in \cite{Dotsenko2013p} and  \cite{PS2011}
one has to linearize the factor $\mathrm{e}^{-xml}$, see (\ref{lin}). \cite{Dotsenko2013p} uses a threefold Gaussian integration, whereas 
\cite{PS2011} uses the differential operator of (\ref{lin}). If already in  \cite{Dotsenko2013p} Eq.~(34) one would use the latter method, then $N_\infty$ would appear without further efforts.
\section{A simplified proof for the Fredholm determinant of $N$ to determine the Airy$_2$ process}
In \cite{PS2011} it is proved that the Fredholm determinant of $N$ 
equals the one of the two-point distribution of the Airy$_2$ process as $t\to\i$. The arguments of  \cite{PS2011}
are somewhat involved. Here we provide a more straightforward proof. 

We start from an expression $G_{t,x_1,x_2}^{\#}(s_1-x_1^2/2t,s_2-x_2^2/2t)=\det(1-\tilde{N})$, which follows from 
multiplying out the last factor of (\ref{Gfact}), with the kernel of $\tilde{N}$ given by  
\begin{eqnarray}
&&\hspace{-30pt}\tilde{N}(z,z') 
=
\mathds{1}_{\{z,z'>0\}}\sum_{n=0}^\i \sum_{k=0}^n (-1)^n \binom{n}{k} 
\mathrm{e}^{-(s_1-x_1^2/2t)k-(s_2- x_2^2/2t)(n-k)} \notag\\
&&\hspace{24pt}  \times \mathrm{e}^{-\frac12 x k(n-k)}
\int du \Ai(u+z+y) \Ai(u+z'-y) \mathrm{e}^{\gamma_t u n}
\end{eqnarray}
with $y=(x_1 k+x_2(n-k))/2\gamma_t$. 
Let us rewrite the integral over $u$ as 
\begin{eqnarray}
 &&\hspace{-35pt}\int du dv \delta(u-v) \mathrm{e}^{\gamma_t un} \Ai(u+z+y) \Ai(v+z'-y) \notag\\
 &&\hspace{-30pt}=
 \int du dv  \delta(u-v) \mathrm{e}^{\gamma_t n w} 
 \mathrm{e}^{\gamma_t u(n-k)}  \Ai \big( u+z+(n-k)x/2\gamma_t\big)
 \mathrm{e}^{\gamma_t v k}  \Ai \big( v+z'+ k x/2\gamma_t\big) \, , 
\end{eqnarray}
where the integrals over $u,v$ have both been shifted  by 
$w= (-x_1 k+x_2(n-k))/2\gamma_t$. We notice the identity
\begin{equation}
\label{Ai_id}
\mathrm{e}^{a u} \Ai (u+\l) 
=
\mathrm{e}^{-\tau(\l + a^2)} \mathrm{e}^{a \tau^2} \mathrm{e}^{-\tau H} \mathrm{e}^{a u} \Ai(v+\l-2\tau a)
\end{equation}
with the Airy operator $H = - (\partial_u)^2 + u$, see Appendix B of \cite{PS2011}. \medskip\\
{\it Proof.} 
Since
\begin{equation}
 (H+\l) \Ai(u+\l) = 0\,, 
\end{equation}
one arrives at  
\begin{equation}
\mathrm{e}^{a u} \Ai(u+\l) = \mathrm{e}^{-\tau (H+\l)} \mathrm{e}^{\tau (H+\l)}\mathrm{e}^{a u} \mathrm{e}^{-\tau (H+\l)} \Ai(u+\l)\,. 
\end{equation}
Using
\begin{align}
 u(\tau) &= \mathrm{e}^{\tau H} u \mathrm{e}^{-\tau H} = u +\tau^2-2\tau\partial_u \,, \notag\\
 \mathrm{e}^{au\tau} 
 &= 
 \mathrm{e}^{a\tau^2} \mathrm{e}^{au} \mathrm{e}^{-2a\tau\partial_u} e^{-\frac12 [au,-2a\tau\partial_u ]}
 =
 \mathrm{e}^{a\tau^2-a^2\tau} \mathrm{e}^{au} \mathrm{e}^{-2a\tau\partial_u}, 
\end{align}
(\ref{Ai_id}) follows from
\begin{equation}
 \mathrm{e}^{a u} \Ai(u+\l) 
 =
 \mathrm{e}^{-\tau (\l + a^2)} \mathrm{e}^{a \tau^2} \mathrm{e}^{-\tau H} \mathrm{e}^{a u} \mathrm{e}^{-2\tau a \partial_u} \Ai(u+\l)\,.
\end{equation}\qed\medskip\\
With the help of identity (\ref{Ai_id})  the kernel of $\tilde{N}$ can be rewritten as 
\begin{eqnarray}
&&\hspace{-60pt}\tilde{N}(z,z')
=
\mathds{1}_{\{z,z'>0\}} \sum_{n=0}^\i \sum_{k=0}^n (-1)^{n-1} \binom{n}{k} 
\mathrm{e}^{-s_1 k- s_2 (n-k)} \mathrm{e}^{(x_1z' - x_2 z)/2\gamma_t^2 }  \notag\\
&&\hspace{-8pt}\times 
\int du dv \langle u | \mathrm{e}^{- xH/2\gamma_t^2} |v\rangle 
\mathrm{e}^{\gamma_t (n-k)u+\gamma_t k v} \Ai(u+z)\Ai(v+z') \,.
\end{eqnarray}
Using the cyclicity of the determinant, $\det(1-AB)=\det(1-BA)$, it holds that
$\det(1- \tilde{N}) = \det(1- \tilde{L})$ with the kernel of $\tilde{L}$ given by
\begin{eqnarray}
&&\hspace{-16pt}\tilde{L}(u,v)
=
\sum_{n=0}^\i \sum_{k=0}^n (-1)^{n-1} \binom{n}{k} 
\mathrm{e}^{-s_1 k- s_2 (n-k)} \langle u| \mathrm{e}^{-x H/ 2\gamma_t^2} |v\rangle \notag\\
&&\hspace{32pt} \times 
\mathrm{e}^{\gamma_t(n-k)u+\gamma_t k v} \int_0^\i \Ai(u+z)\Ai(v+z) \mathrm{e}^{- xz/2\gamma_t^2 } dz\,.  
\end{eqnarray}
From this expression it is easy to see that the Fredholm determinant of $\tilde{L}$
yields the one of  the Airy$_2$ process for $t\to\i$, see the arguments in 
Section 2.4 of \cite{PS2011}. \\\\
\textbf{Acknowledgement}. We are grateful to Victor Dotsenko for useful comments and for communicating his preprint \cite{Dotsenko2013p}.


\end{document}